\documentclass[journal,11pt]{IEEEtran}

\usepackage{amsthm}
\usepackage{amsfonts}
\usepackage{amssymb}
\usepackage{mathrsfs}
\usepackage{mathtools}
\usepackage{float}
\usepackage[pdftex]{graphicx}
\usepackage{amsmath}
\usepackage{multirow,multicol}
\usepackage{graphicx}

\usepackage{times}
\usepackage{textcomp}

\usepackage{verbatim}
\usepackage{balance}
\usepackage[inline]{enumitem}

\usepackage[caption=false,font=footnotesize]{subfig}
\usepackage{cite}
\usepackage{hyperref}

\usepackage[table]{xcolor}
\usepackage{tikz-cd}

\usepackage{tikz}
\usetikzlibrary{decorations.markings}
\usetikzlibrary{decorations.pathmorphing}
\usetikzlibrary{arrows, snakes}
\usetikzlibrary{shapes.geometric, calc,decorations.pathreplacing}
\usetikzlibrary{backgrounds}
\usetikzlibrary{fit}
\usetikzlibrary{decorations.pathreplacing}
\usetikzlibrary{decorations.pathreplacing,decorations.markings}
\input xy
\xyoption{all}

\tikzset{middlearrow/.style={
		decoration={markings,
			mark= at position 0.5 with {\arrow{#1}} ,
		},
		postaction={decorate}
	}
}

\tikzset{
	on each segment/.style={
		decorate,
		decoration={
			show path construction,
			moveto code={},
			lineto code={
				\path [#1]
				(\tikzinputsegmentfirst) -- (\tikzinputsegmentlast);
			},
			curveto code={
				\path [#1] (\tikzinputsegmentfirst)
				.. controls
				(\tikzinputsegmentsupporta) and (\tikzinputsegmentsupportb)
				..
				(\tikzinputsegmentlast);
			},
			closepath code={
				\path [#1]
				(\tikzinputsegmentfirst) -- (\tikzinputsegmentlast);
			},
		},
	},
	mid arrow/.style={postaction={decorate,decoration={
				markings,
				mark=at position .5 with {\arrow[#1]{stealth}}
	}}},
}

\tikzset{
	on each segment/.style={
		decorate,
		decoration={
			show path construction,
			moveto code={},
			lineto code={
				\path [#1]
				(\tikzinputsegmentfirst) -- (\tikzinputsegmentlast);
			},
			curveto code={
				\path [#1] (\tikzinputsegmentfirst)
				.. controls
				(\tikzinputsegmentsupporta) and (\tikzinputsegmentsupportb)
				..
				(\tikzinputsegmentlast);
			},
			closepath code={
				\path [#1]
				(\tikzinputsegmentfirst) -- (\tikzinputsegmentlast);
			},
		},
	},
	end arrow/.style={postaction={decorate,decoration={
				markings,
				mark=at position 1 with {\arrow[#1]{stealth}}
	}}},
}

\newcommand{\dd}{\mathrm{d}}

\DeclareMathOperator*{\argmin}{arg\,min}

\begin{document}
	
	\title{\huge{Information Geometry for the Working Information Theorist}}
	
	\author{ 
			Kumar Vijay Mishra~\IEEEmembership{Senior Member,~IEEE}, M. Ashok Kumar~\IEEEmembership{Member,~IEEE}, Ting-Kam Leonard Wong\\[2ex]
            \textit{Dedicated to the memory of Prof. Calyampudi Radhakrishna Rao (1920-2023)}
            \thanks{K.~V.~M. is with the United States DEVCOM Army Research Laboratory, Adelphi, MD 20783 USA. E-mail: kvm@ieee.org.}
            \thanks{M.~A.~K. is with the Department of Mathematics, Indian Institute of Technology Palakkad, 678557 India. E-mail: ashokm@iitpkd.ac.in.}
            \thanks{T.-K.~L.~W. is with the Department of Statistical Sciences, University of Toronto, Toronto, ON M5S 3G3, Canada. E-mail: tkl.wong@utoronto.ca.}\vspace{-20pt}
	}

	\maketitle
	

\begin{abstract}
Information geometry is a study of statistical manifolds, that is, spaces of probability distributions from a geometric perspective. Its classical information-theoretic applications relate to statistical concepts such as Fisher information, sufficient statistics, and efficient estimators. Today, information geometry has emerged as an interdisciplinary field that finds applications in diverse areas such as radar sensing, array signal processing, quantum physics, deep learning, and optimal transport. This article presents an overview of essential information geometry to initiate an information theorist, who may be unfamiliar with this exciting area of research. We explain the concepts of divergences on statistical manifolds, generalized notions of distances, orthogonality, and geodesics, thereby paving the way for concrete applications and novel theoretical investigations. We also highlight some recent information-geometric developments, which are of interest to the broader information theory community.
\end{abstract}
\vspace{-10pt}
\section{Introduction}
{\it Information geometry} provides geometric concepts and tools for studying spaces of probability distributions. These spaces arise naturally in theory and application. For example, the unit simplex describes possible distributions of independent signals from a finite alphabet. Statistical models, such as the multivariate normal distribution, can be regarded as collections of probability distributions on appropriate state spaces. The earliest information geometric ideas are traced to C. R. Rao's seminal 1945 paper \cite[Secs.~6,7]{rao1945information}, a contribution that also led him to recently win the International Statistics Prize in 2023. 

Remarkably, several information-theoretic problems including parameter estimation, entropy maximization, dimension reduction, and interpolation between distributions can be interpreted from a geometric point of view. For example, the information geometry bases the distance between the parameterized probability distributions on the Fisher information matrix (FIM) \cite{amari2000methods}, which is the inverse of the deterministic {\it Cram\'{e}r-Rao lower bound (CRLB)}. Therefore, the results derived from information geometry are directly connected with the fundamentals of estimation theory. These kinds of geometric approaches give valuable insights into designing and understanding algorithms leading to interesting and foundational mathematical questions. 
To quote Amari and Nagaoka \cite{amari2000methods}:
\begin{quote}
{\it Information geometry ... allows us to take problems from a variety of fields: statistics, information theory, and control theory; visualize them geometrically; and from this develop novel tools with which to extend and advance these fields.}
\end{quote}   
We may now extend the above list of applications (written more than two decades ago) to include quantum computing, mathematical biology, machine learning, radar remote sensing, deep learning, array signal processing, quantitative finance, statistical physics, image processing, and covariant thermodynamics (see, e.g., \cite{amari2000methods, amari2016information, plastino2021information}, and references therein). Very recently, information geometry was used to analyze Higgs boson measurements at the Large Hadron Collider \cite{brehmer2017better}. At the time of writing this article, a search on Google Scholar returns more than six thousand papers associated with the keyword ``information geometry" since 2019. 
The aim of this article, whose title is inspired from \cite{mac2013categories}, is to provide a modern and accessible exposition of the key ideas of information geometry, and describe some recent developments which are of interest to researchers in information theory and related fields. 

The term information geometry is customarily associated with the study of statistical models using differential geometry, especially Fisher-Rao geometry and an extension called {\it dualistic geometry}. 
In this article, we advocate {\it information geometry in the broad sense} -- the geometric mindset rather than specific frameworks. There are different geometric structures on spaces of probability distributions, and the most appropriate one depends on the problem in hand. 
Throughout this article, we use {\it divergence} -- a notion of generalized distance -- as a unifying theme. This allows us to explain the main ideas without explicit use of differential geometric notations. This article should be viewed as complementing other recent surveys \cite{Amari2021Information, nielsen2020elementary, nielsen2022many}, which present information geometry from a detailed mathematical perspective using the language of differential geometry. For enhanced readability, we omit detailed theoretical guarantees/conditions and refer the reader to the corresponding references. There exist other approaches such as {\it algebraic statistics}, wherein algebraic geometry is applied to statistical models containing singularities not amenable to differential geometric treatment \cite{sullivant2018algebraic}. The mathematical tools involved here are considerably different and, therefore, we do not discuss them in this article. 

\vspace{-10pt}
\section{Divergence and its geometry} \label{sec:div}
The key idea of information geometry is to regard a space of probability distributions as a geometric object called a {\it statistical manifold}. For example, consider the collection of normal distributions on the real line. Since each element is uniquely characterized by its mean $\mu \in \mathbb{R}$ and standard deviation $\sigma > 0$, we may regard the collection as a two dimensional surface and identify it with the upper half plane. Other parameterizations may be used, e.g., $\theta_1 = \frac{\mu}{\sigma^2}$ and $\theta_2 = \frac{-1}{2\sigma^2}$ are the so-called canonical parameters when we view the normal distribution as an {\it exponential family}. 
Depending on the application, we may consider high dimensional statistical manifolds (such as neural networks \cite{amari1992information}) or even infinite dimensional ones (such as the space $\mathcal{P}([a, b])$ of all probability distributions on an interval). 
\vspace{-10pt}
\subsection{Divergence} \label{sec:divergence}
Denote a statistical manifold by $\mathcal{M}$, where each element $\mathbb{P}$ is a probability distribution on some state space $\mathcal{X}$, say a finite alphabet or a domain in $\mathbb{R}^n$. The beginning of a geometric structure on $\mathcal{M}$ is a notion of distance between elements $\mathbb{P}$, $\mathbb{Q}$ of $\mathcal{M}$. In contrast to ordinary geometry, here it is crucial to allow generalized distances which are possibly {\it asymmetric} in the two arguments. Hence they are generally not metrics and do not satisfy the triangle inequality. Formally, a {\it divergence} defines for each pair $\mathbb{P}, \mathbb{Q}$ a directed distance ${\bf D}(\mathbb{P} || \mathbb{Q}) \geq 0$, such that ${\bf D}(\mathbb{P} || \mathbb{Q}) = 0$ if and only if $\mathbb{P} = \mathbb{Q}$. 
Even without additional geometric structures, a divergence is already useful for e.g.~serving as a loss function for model fitting (as in the very influential {\it Wasserstein  generative adversarial networks} (GANs) \cite{arjovsky2017wasserstein}) and constructing distributional neighborhoods for quantifying model uncertainty \cite{rahimian2019distributionally}. We also mention that the term divergence is also used to mean a generalized distance on a generic set, which is not necessarily a statistical manifold. 

One of the most familiar and important example is the {\it Kullback-Leibler (KL) divergence} which plays a fundamental role in information theory, statistics and related fields:
\begin{equation} \label{eqn:KL.divergence}
{\bf H}(\mathbb{P} || \mathbb{Q}) = \int_{\mathcal{X}} \log \frac{\dd \mathbb{P}}{\dd \mathbb{Q}} \dd \mathbb{P}.
\end{equation}
The KL-divergence illustrates a general recipe for creating divergences between probability distributions. Namely, compare $\mathbb{P}$ and $\mathbb{Q}$ {\it pointwise} via the Radon-Nikodym derivative (or {\it likelihood ratio}) $\frac{\dd \mathbb{P}}{\dd \mathbb{Q}}$, and then form a weighted average. Two explicit examples of this exercise are the {\it $f$-divergence}
\begin{equation} \label{eqn:f.divergence}
\int_{\mathcal{X}} f\Big( \frac{\dd \mathbb{P}}{\dd \mathbb{Q}} \Big) \dd \mathbb{Q}, \quad \text{($f$ convex, $f(1) = 0$)},
\end{equation}
and the {\it R\'{e}nyi divergence}
\begin{equation} \label{eqn:Renyi.divergence}
\frac{1}{\alpha - 1} \log \int_{\mathcal{X}} \Big( \frac{\dd \mathbb{P}}{\dd \mathbb{Q}} \Big)^{\alpha} \dd \mathbb{Q}, \quad (\alpha > 0,  \neq 1),
\end{equation}
both of which generalize the KL-divergence.

The above divergences, which are constructed in terms of pointwise comparison of the likelihood ratio, satisfy {\it information monotonicity}. That is, if $\tilde{\mathbb{P}}$ and $\tilde{\mathbb{Q}}$ are transformations of $\mathbb{P}$ and $\mathbb{Q}$ via stochastic channels, then ${\bf D}(\tilde{\mathbb{P}} || \tilde{\mathbb{Q}}) \leq {\bf D}( \mathbb{P} || \mathbb{Q})$. This is closely related to the {\it data-processing inequality} in information theory. A detailed study of this inequality when $\mathbb{P} \approx \mathbb{Q}$, as well as the case of equality ({\it invariance} under sufficient statistics), leads to the {\it Fisher-Rao information metric} \eqref{eqn:Hessian.as.Fisher.Rao} whose geometry plays a fundamental role in information geometry and its applications.

Divergences based on the likelihood ratio are sometimes too restrictive in modern applications. To give a simple example, consider the normal distributions $\mathbb{P} = N(0, \epsilon^2)$ and $\mathbb{Q} = N(\sqrt{\epsilon}, \epsilon^2)$. Then ${\bf H}(\mathbb{P} || \mathbb{Q}) = \frac{1}{2\epsilon}$ diverges to infinity as $\epsilon \downarrow 0$, even though both distributions are effectively the point mass at zero when $\epsilon$ is small. The problem is that the distributions are almost supported on disjoint sets (note that $\epsilon \ll \sqrt{\epsilon}$ as $\epsilon \downarrow 0$). When this happens, likelihood ratios are ineffective as they neglect the geometry of the underlying state space. 

This leads us to another general recipe, mainly based on {\it optimal transport}, for constructing divergences. 
Consider probability distributions on a state space $\mathcal{X}$. Equip the state space $\mathcal{X}$ with its own geometry via a divergence $D(\cdot||\cdot)$. That is, we have a generalized distance $D(x||y) \geq 0$ for points $x, y \in \mathcal{X}$. Identify the points $x, y$ with point masses, i.e., $x \leftrightarrow \mathbb{P} = \delta_x$ and $y \leftrightarrow \mathbb{Q} = \delta_y$, and define ${\bf D}(\delta_x || \delta_y) = D(x||y)$. Our goal is to {\it lift} the divergence from $\mathcal{X}$ to construct a divergence ${\bf D} (\mathbb{P} || \mathbb{Q})$ for $\mathbb{P}, \mathbb{Q} \in \mathcal{P}(\mathcal{X})$, the space of probability distributions on $\mathcal{X}$. Optimal transport achieves this by defining
\begin{equation} \label{eqn:Kantorovich}
{\bf D}(\mathbb{P} || \mathbb{Q}) = \inf_{(X, Y): X \sim \mathbb{P}, Y \sim \mathbb{Q}} \mathbb{E} D(X||Y),
\end{equation}
where the infimum is over pairs $(X, Y)$ of $\mathcal{X}$-valued random variables satisfying $X \sim \mathbb{P}$ and $Y \sim \mathbb{Q}$. In probability theory such a pair (or rather its joint distribution) is called a {\it coupling} of $\mathbb{P}$ and $\mathbb{Q}$. Intuitively, \eqref{eqn:Kantorovich} compares $\mathbb{P}$ and $\mathbb{Q}$ ``horizontally'', rather than ``vertically'' as in \eqref{eqn:KL.divergence}.

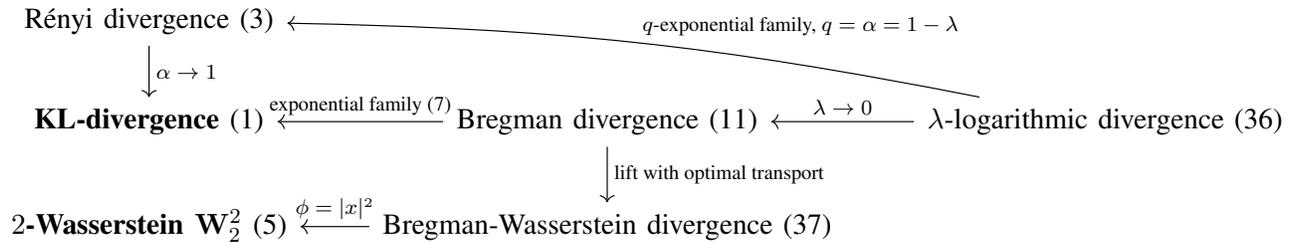
\begin{figure*}[t!]
\centering
\begin{tikzcd}
\text{R\'{e}nyi divergence } \eqref{eqn:Renyi.divergence} \arrow[d, "\text{$\alpha \rightarrow 1$}"] \arrow[drr, leftarrow, bend left = 5, "\text{$q$-exponential family, $q = \alpha = 1 - \lambda$}"] & & \\
\textbf{KL-divergence } \eqref{eqn:KL.divergence}  \arrow[r, leftarrow, "\text{exponential family \eqref{eqn:exp.family}}"] &
\text{Bregman divergence } \eqref{eqn:Bregman} \arrow[r, leftarrow, "\text{$\lambda \rightarrow 0$}"] \arrow[d, "\text{lift with optimal transport}"] & \text{$\lambda$-logarithmic divergence }  \eqref{eqn:log.divergence} \\
\textbf{$2$-Wasserstein ${\bf W}_2^2$ } \eqref{eqn:W2} \arrow[r, leftarrow, "\text{$\phi = |x|^2$}"] & \text{Bregman-Wasserstein divergence } \eqref{eqn:BW} 
\end{tikzcd}
\caption{Relations among some of the divergences discussed in this article. The KL-divergence and the (squared) $2$-Wasserstein distance, which are shown in bold font, are basic examples of the two major approaches (likelihood ratio and optimal transport) for constructing divergences between probability distributions.} \label{fig:summary}
\end{figure*}

We illustrate these concepts through a few examples. First, assume $D(x||y) = 1_{x \neq y}$ is the discrete metric. From \eqref{eqn:Kantorovich}, we want to couple $X$ and $Y$ to maximize $\mathbb{P}(X = Y)$, a task common in information theory. The resulting divergence is
\[
{\bf D}(\mathbb{P}||\mathbb{Q}) = \frac{1}{2} \| \mathbb{P} - \mathbb{Q} \|_{TV},
\]
which is a multiple of the {\it total variation distance}. Next suppose $\mathcal{X} = \mathbb{R}^n$ and let $D(x||y) = |x - y|^2$ be the squared Euclidean distance. Then
\begin{equation} \label{eqn:W2}
{\bf W}_2(\mathbb{P}, \mathbb{Q}) := {\bf D}(\mathbb{P} || \mathbb{Q})^{1/2}
\end{equation}
is the celebrated {\it $2$-Wasserstein distance} on $\mathcal{P}(\mathbb{R}^d)$ (here $\mathcal{X}$ can be replaced by any metric space). Suppose we compare $\mathbb{P} = N(0, \epsilon^2)$ and $\mathbb{Q} = N(\sqrt{\epsilon}, \epsilon^2)$ using the $2$-Wasserstein distance. Using the translation $Y = X + \sqrt{\epsilon}$, it can be shown that ${\bf W}_2(\mathbb{P}, \mathbb{Q}) = \sqrt{\epsilon}$. Thus optimal transport can be used to construct divergences which are continuous with respect to spatial perturbations of distributions. By extending the dynamic formulation \eqref{eqn:W2.dynamic.cost} of optimal transport, divergences that interpolate between the Wasserstein and Fisher-Rao (Riemannian) distances have recently been introduced \cite{chizat2018interpolating}. The relations among some divergences discussed in this article are illustrated in Figure \ref{fig:summary}.
\vspace{-10pt}
\subsection{Geometry induced by a divergence} \label{sec:geometry.divergence}
A divergence ${\bf D}(\mathbb{P}|| \mathbb{Q})$ defines a generalized distance between $\mathbb{P}$, $\mathbb{Q}$ in some statistical manifold $\mathcal{M} \subset \mathcal{P}(\mathcal{X})$. Under suitable conditions, the divergence can be used to define the following additional geometric concepts: 
\begin{itemize}
    \item (Speed and orthogonality) Suppose $(\mathbb{P}_t)_{t \geq 0}$ is a distribution-valued path. We want to speak of the {\it instantaneous speed} of the path at each time $t$. Also, given paths $(\mathbb{P}_t)_{t \geq 0}$ and $(\mathbb{Q}_t)_{t \geq 0}$ starting at the same point, we want to be able to say whether these two paths meet {\it orthogonally} at $t = 0$. These correspond to the concept of {\it Riemannian metric}.
    \item (Geodesic) Given distributions $\mathbb{P}_0$ and $\mathbb{P}_1$, we want to construct {\it interpolating paths} $(\mathbb{P}_t)_{0 \leq t \leq 1}$ which are in some sense most natural. They can be defined by a distance minimizing property or by enforcing the ``acceleration'' of the path to be everywhere zero.  
\end{itemize}

One application in information theory, where all three geometric notions (divergence, orthogonality, and geodesic) combine naturally, is {\it information projection} studied by Csisz\'{a}r and others \cite{csiszar1984sanov, kumar2015minimization-1, kumar2015minimization-2}. In abstract terms, let $\mathcal{S} \subset \mathcal{M}$ be a subspace of $\mathcal{M}$ and let $\mathbb{P} \in \mathcal{M}$. Given a divergence ${\bf D}(\cdot||\cdot)$, the problem is to find $\mathbb{Q}^* \in \mathcal{S}$ which is {\it closest} to $\mathbb{P}$. Since the divergence may be asymmetric, there are generally {\it two} ways to formulate the projection:
\begin{equation} \label{eqn:I.projection}
 \mathbb{Q}^* = \left\{\begin{array}{l}
        \argmin_{\mathbb{Q} \in \mathcal{S}} {\bf D}(\mathbb{Q} || \mathbb{P}), \text{ or}\\
        \argmin_{\mathbb{Q} \in \mathcal{S}} {\bf D}(\mathbb{P} || \mathbb{Q}). \\
        \end{array}\right.
\end{equation}
When ${\bf D}(\cdot||\cdot)$ is the KL-divergence, information projection is closely related to diverse topics including Sanov's theorem, maximum likelihood estimation, maximum entropy principle, and Schr\"{o}dinger bridge in optimal transport. Under suitable conditions, the projection $\mathbb{Q}^*$ can be characterized by the geometric property that an appropriately defined geodesic from $\mathbb{P}$ to $\mathbb{Q}^*$ is orthogonal to $\mathcal{S}$. Since there are two versions of projections, in general, there should be two versions of geodesic. This leads us to the {\it dualistic geometry}, which is an essential feature of classical information geometry. 

\vspace{-10pt}
\subsection{Applications}
Methodologies based on divergence geometry and minimization have flourished in the last few decades; see, e.g., \cite{eguchi2022minimum} for a modern treatment with a focus on statistics and machine learning. Because of its analytical and computational tractability, as well as connections with the exponential family, the {\it Bregman divergence} \eqref{eqn:Bregman}, and its variations, are especially useful in applications including clustering, nonlinear principle component analysis, boosting, and optimization \cite{amari2016information}. In particular, we highlight the {\it mirror descent} \cite{beck2003mirror} which is a non-Euclidean gradient descent algorithm and has been extended in many directions to solve large-scale deterministic and stochastic optimization problems. 
\vspace{-10pt}
\section{Dualistic geometry} \label{sec:exp.family}
{\it Dualistic geometry} is an extension of Riemannian geometry which was first discovered in \cite{amari1982differential1} and applied afterward to study asymptotic properties in statistical inference. In a nutshell, dualistic geometry uses a divergence to construct the structures mentioned in Section \ref{sec:geometry.divergence}: notions of speed, orthogonality, and geodesics which are {\it compatible} (this is what duality means). While the precise formulation of dualistic geometry requires some terminologies from differential geometry, we will explain the main ideas with a concrete example: a statistical manifold consisting of probability distributions that form an {\it exponential family}. 
\vspace{-10pt}
\subsection{Exponential family}
Let the state space $\mathcal{X}$ be equipped with a reference measure $\nu$, such as the Lebesgue measure on $\mathbb{R}^d$ or the counting measure on a discrete set. A statistical manifold $\mathcal{M} = \{ \mathbb{P}_{\theta} \}_{\theta \in \Theta} \subset \mathcal{P}(\mathcal{X})$, parameterized by $\theta \in \Theta \subset \mathbb{R}^d$ (here $\Theta$ is an open convex set), is an {\it exponential family} if each $\mathbb{P}_{\theta}$ has a  density with respect to $\nu$ of the form
\begin{equation} \label{eqn:exp.family}
p_{\theta}(x) = \frac{\dd \mathbb{P}_{\theta}}{\dd \nu}(x) = e^{\theta \cdot F(x) - \phi(\theta)},
\end{equation}
where $\theta = (\theta_1, \ldots, \theta_d)$ is the {\it canonical parameter}, $F(x) = (F_1(x), \ldots, F_d(x)) : \mathcal{X} \rightarrow \mathbb{R}^d$ is a vector of statistics, and $\phi(\theta)$ is a normalization factor which can be shown to be a {\it convex} function of $\theta$. For example, the density of the normal distribution $N(\mu, \sigma^2)$ can be expressed as an exponential family where $\nu$ is the Lebesgue measure, $d = 2$ and $F(x) = (x, x^2)$. 

Exponential families can be motivated by  {\it entropy maximization} which is fundamental in information theory \cite{cover2006elements}. Given a density $p$ with respect to $\nu$, define its {\it differential entropy} by
\begin{equation} \label{eqn:differential.entropy}
{\bf H}(p) = -\int_{\mathcal{X}} p(x) \log p(x) \dd \nu(x).
\end{equation}
Consider the entropy maximization problem
\begin{equation} \label{eqn:max.ent}
\sup_{p: p > 0} {\bf H}(p) \text{ subject to } \mathbb{E}_{p}[F(X)] = c,
\end{equation}
where $X \sim p$ and $c \in \mathbb{R}^d$ is a given constant. Then, it can be shown that the optimal density $p$, when exists, has the form \eqref{eqn:exp.family}. For example, if $\nu$ is the Lebesgue measure on $\mathbb{R}$ and $F(x) = (x, x^2)$, we recover the well known result that the normal distribution maximizes the differential entropy when the first two moments are fixed. In fact, the concept of exponential family originated from the {\it Boltzmann-Gibbs distributions} in statistical physics; they are equilibrium distributions of physical systems in which entropy is nondecreasing. Note that when $\nu$ itself is a probability measure then $-{\bf H}(p) = {\bf H}(\mathbb{P} || \nu)$ is the KL-divergence (where $\dd \mathbb{P} = p \dd \nu$). So \eqref{eqn:max.ent} is equivalent to the information projection problem $\inf_{ \mathbb{P} \in \mathcal{S} } {\bf H}( \mathbb{P} || \nu)$ as in \eqref{eqn:I.projection}, where the set $\mathcal{S}$ captures the relevant constraints. 

Due to the success of the maximum entropy principle, there have been numerous attempts to construct {\it generalized exponential families} by using generalized entropies and other constraints. In particular, using the {\it Tsallis} or {\it R\'{e}nyi entropy} (and using the so-called {\it escort expectation}) leads to the {\it $q$-exponential family} \cite{amari2011geometry, WZ21} which models power-law behaviours in {\it nonextensive statistical physics}. Generalized exponential families have also found applications in machine learning such as clustering and attention mechanisms, see e.g. \cite{amid2023clustering, martins2022sparse}.

\vspace{-10pt}
\subsection{Bregman divergence and its geometry}
\label{subsec:bregman}
To develop a geometric structure on the exponential family $\mathcal{M}$, consider the KL-divergence ${\bf H}(\mathbb{P}_{\theta'} || \mathbb{P}_{\theta})$ which, after restricting to $\mathcal{M}$, can be regarded as a function of $(\theta, \theta') \in \Theta \times \Theta$. Using \eqref{eqn:KL.divergence} and \eqref{eqn:exp.family}, one can show that
\begin{equation} \label{eqn:KL.as.Bregman}
{\bf H}(\mathbb{P}_{\theta'} || \mathbb{P}_{\theta}) = D_{\phi}(\theta || \theta'),
\end{equation}
where
\begin{equation} \label{eqn:Bregman}
D_{\phi}(\theta || \theta') = \phi(\theta) - \phi(\theta') - \nabla \phi(\theta') \cdot (\theta - \theta')
\end{equation}
is the {\it Bregman divergence} induced by the convex normalizing function $\phi(\theta)$. One should think of the Bregman divergence as a non-Euclidean generalization of the {\it square loss} $\frac{1}{2}|\theta - \theta'|^2$ (recovered when $\phi(\theta) = \frac{1}{2}|\theta|^2$). Thus the information geometry of exponential family boils down to the information geometry of Bregman divergence.  Convex duality, which is not present in the general case, makes the Bregman divergence and its geometry particularly tractable and hence useful in applications. 

Consider the behaviour of $D_{\phi}(\theta + \Delta \theta || \theta)$ for a small displacement $\Delta \theta$. The second order Taylor approximation gives
\[
D_{\phi}(\theta + \Delta \theta || \theta) \approx \frac{1}{2} \sum_{i, j} \partial_{ij} \phi(\theta) \Delta \theta_i \Delta \theta_j.
\]
The right hand side, which involves the Hessian of $\phi$, defines the {\it norm} $\|v\|$ of a displacement $v$ started at $\theta$:
\begin{equation} \label{eqn:Bregman.metric}
\|v\|_{\theta}^2 = \sum_{i, j} \partial_{ij} \phi(\theta) v_i v_j.
\end{equation}
We get the same norm if we expand $D_{\phi}(\theta || \theta + \Delta \theta )$ instead. By polarization, two directions $u, v$ are said to be {\it orthogonal} at $\theta$ if
\begin{equation} \label{eqn:Bregman.inner.product}
\langle u, v \rangle_{\theta} = \sum_{i, j} \partial_{ij} \phi(\theta) u_i v_j = 0.
\end{equation}
In our setting, the Hessian of $\phi$ can be interpreted probabilistically as the {\it Fisher information matrix}: $\partial_{ij}\phi(\theta) = g_{ij}(\theta)$, where
\begin{equation} \label{eqn:Hessian.as.Fisher.Rao}
g_{ij}(\theta) = \mathbb{E}_{\theta} \left[ \partial_i \log p_{\theta}(X) \partial_i \log p_{\theta}(X) \right].
\end{equation}
As observed by Rao \cite{rao1945information} in 1945, the matrix $G(\theta) = (g_{ij}(\theta))_{i, j}$, which can be defined for any (sufficiently regular) parameterized density, indeed defines a Riemannian metric, called the {\it Fisher-Rao metric} because it transforms appropriately under reparameterizations, i.e., all parameterizations define the same quantity. It is a general fact that the KL-divergence between parameterized densities can be locally approximated by the Fisher-Rao metric: 
\begin{equation} \label{eqn:KL.Hessian}
g_{ij}(\theta) = \left. \frac{\partial^2}{\partial \theta_i' \partial \theta_j'} {\bf H}( \mathbb{P}_{\theta} || \mathbb{P}_{\theta'}) \right|_{\theta' = \theta}.
\end{equation}
In particular, the Fisher-Rao metric characterizes the first order conditions in the projection problem \eqref{eqn:I.projection} when the underlying divergence is the KL-divergence. 

For the Bregman divergence, geodesics can be defined in terms of {\it convex duality} \cite{rockafellar1997convex}. Recall the {\it convex conjugate} $\phi^*(\eta) = \sup_{\theta} \left\{ \theta \cdot \eta - \phi(\theta) \right\}$ which defines a convex function in $\eta$. The fundamental {\it Fenchel-Young inequality}
states that, for any $\theta$ and $\eta$, we have
\begin{equation} \label{eqn:Fenchel.Young}
\phi(\theta) + \phi^*(\eta) \geq \theta \cdot \eta,
\end{equation}
and equality holds if and only if $\eta = \nabla \phi(\theta)$. This explains the fundamental role played by the {\it Legendre transformation} $\eta = \nabla \phi(\theta)$, which is also called the {\it mirror map} in machine learning applications involving the Bregman divergence. Its inverse is $\theta = \nabla \phi^*(\eta)$. Geometrically, $\eta = \nabla \phi(\theta)$ defines an alternative coordinate system. For an exponential family, we have the probabilistic interpretation $\eta = \mathbb{E}_{\theta} F(X)$. Hence we call it the {\it expectation parameter}.

\begin{figure}[t!]
	\centering
	\vspace{-0.5cm}
	\includegraphics[scale = 0.4]{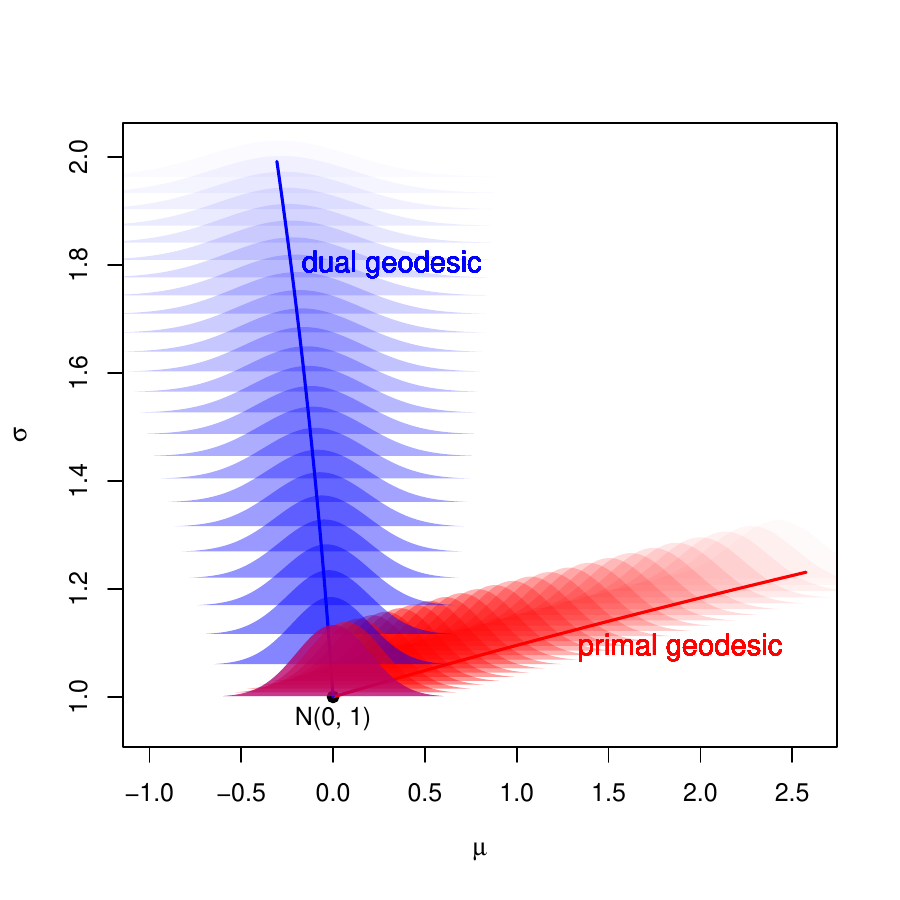}
    \vspace{-0.5cm}
	\caption{A primal geodesic (in red) and a dual geodesic (in blue), in the space of univariate normal distribution, both started at $N(0, 1)$. Note that both are not straight lines when visualized on the $(\mu, \sigma)$-halfplane as in the figure. We also visualize the corresponding densities (not in scale). These two geodesics are constructed to be orthogonal at $N(0, 1)$ with respect to the Fisher-Rao metric.} \label{fig:normal_geodesic}
\end{figure}

Consider a curve $(\mathbb{P}_t)$ in $\mathcal{M}$, where $t$ stands for time. Each $\mathbb{P}_t$ corresponds to a canonical parameter $\theta_t$ and an expectation parameter $\eta_t$ related by $\eta_t = \nabla \phi(\theta_t)$. We say that $\mathbb{P}_t$ is:
\begin{itemize}
\item a {\it primal geodesic}, if $(\theta_t)$ is a constant velocity straight line;
\item a {\it dual geodesic}, if $(\eta_t)$ is a constant velocity straight line. 
\end{itemize}
See Figure \ref{fig:normal_geodesic} for an illustration within the family of normal distributions on $\mathbb{R}$.

These geodesics provide two ways of interpolating between elements $\mathbb{P}_{\theta_0}$ and $\mathbb{P}_{\theta_1}$ of $\mathcal{M}$. The primal geodesic yields the {\it exponential interpolation}: $\theta_t = (1 - t) \theta_0 + t \theta_1$. In terms of the density, we have $p_t(x) = \frac{1}{Z_t} p_0(x)^{1 - t} p_1(x)^t$, where $Z_t$ is a normalizing factor. This interpolation features in e.g.~{\it annealed importance sampling} \cite{neal2001annealed}. Analgously, the dual geodesic yields the {\it mixture interpolation} satisfying $\eta_t = (1 - t) \eta_0 + t \eta_1$. This gives a density of the form $p_t(x) = \frac{1}{Z_t} e^{\nabla \phi^*(\eta_t) \cdot F(x)}$. Note that both geodesics are generally different from the usual {\it mixture density} $(1 - t) p_0(x) + t p_1(x)$. 

The Hessian in \eqref{eqn:Bregman.metric}, as well as the primal and geodesics, are {\it dual} or simply compatible in the sense that the inner product between primal and dual geodesics started at the same point has a simple expression. This duality is used to prove the {\it generalized Pythagorean theorem} stated as follows; while it holds for any Bregman divergence, for concreteness we phrase it in terms of the KL-divergence and the exponential family $\mathcal{M}$. Assume $\mathbb{P}$, $\mathbb{P}'$ and $\mathbb{P}''$ to be elements of $\mathcal{M}$. Then the generalized Pythagorean relation\[
{\bf H}( \mathbb{P}'' || \mathbb{P}') + {\bf H}( \mathbb{P}' || \mathbb{P}) = {\bf H}( \mathbb{P}'' || \mathbb{P})
\]
holds if and only if the primal geodesic from $\mathbb{P}'$ to $\mathbb{P}$ and the dual geodesic from $\mathbb{P}'$ to $\mathbb{P}''$ are orthogonal, with respect to the Fisher-Rao metric, at $\mathbb{P}'$. This gives a geometric characterization of {\it Bregman projections} which is a special case of information projection (see \cite{nielsen2018information}). 

Note that, while the classical Pythagorean theorem is proved on a two-dimensional Euclidean space, its analogue in information geometry replaces the square Euclidean distance by the KL-divergence (KL), with which the pair of primal and dual geodesics is concomitant. An interesting application of this is found in regression analysis, where least-squares estimate (LSE) coincides with maximum likelihood estimation (MLE) when Guassian distribution is assumed. This is achieved by a direct application of Pythagorean theorem to the Euclidean space. Recent studies have exploited the information-geometric generalization of Pythagorean theorem to derive non-MLE estimators that coincide with LSE for non-Gaussian distributions \cite{eguchi2021pythagoras}.

The general theory of dualistic geometry, which can be found in \cite{amari2000methods, amari2016information}, explains how orthogonality and geodesics can be defined for any ``locally quadratic'' divergence on a (typically) finite dimensional statistical manifold. Nevertheless, only special divergences, such as the Bregman divergence and the recently developed {\it logarithmic divergence} \eqref{eqn:log.divergence} \cite{w18, WZ21}, among some others, admit tractable geodesics and the generalized Pythagorean theorem.
We also note that {\it nonparametric information geometry} can be constructed (see e.g. \cite{pistone2013nonparametric}) but typically involves functional analytic techniques such as Orlicz spaces. Developing associated methodologies and algorithms is an interesting direction. 


\section{Fisher-Rao metric and beyond} \label{sec:Fisher.Rao}

Apart from being the second order local approximation of the KL-divergence (see \eqref{eqn:KL.Hessian}), the Fisher-Rao metric \eqref{eqn:Hessian.as.Fisher.Rao} can be understood in terms of {\it information invariance}. Let $\mathcal{M} = \{ \mathbb{P}_{\theta} \}_{\theta \in \Theta}$ be a statistical model for the distribution of a random variable $X$. Let $Y = T(X)$ be a function of $X$. For example, if $X = (X_1, \ldots, X_n)$ we may consider the sample mean $Y = \frac{1}{n}(X_1 + \cdots + X_n)$. The distribution of $Y$ yields an induced statistical model $\mathcal{N} = \{ \mathbb{Q}_{\theta} \}_{\theta \in \Theta}$. The data processing inequality implies that ${\bf H}(\mathbb{Q}_{\theta}|| \mathbb{Q}_{\theta'}) \leq {\bf H}(\mathbb{P}_{\theta}|| \mathbb{P}_{\theta'})$. Letting $\theta' \rightarrow \theta$ leads to the inequality $G^{\mathcal{N}}(\theta) \leq G^{\mathcal{M}}(\theta)$ for the corresponding Fisher information matrices (this means that $G^{\mathcal{M}}(\theta) - G^{\mathcal{N}}(\theta)$ is positive semidefinite). We say that $Y = T(X)$ is {\it sufficient} for $\mathcal{M}$ if the conditional distribution of $Y$ given $X$ is independent of $\theta$, i.e., $\theta \rightarrow Y = T(X) \rightarrow X$ forms a Markov chain. Intuitively, this means that all information about $\theta$ is contained in $Y$. For example, for an exponential family \eqref{eqn:exp.family} we have that $Y = F(X)$ is sufficient. When $Y$ is sufficient, it can be shown that $G^{\mathcal{N}}(\theta) = G^{\mathcal{M}}(\theta)$, i.e., the Fisher-Rao metric is invariant. {\it Chentsov's theorem} states that the Fisher-Rao metric is the unique Riemannnian metric (up to a multiplicative constant) with this property. See \cite{fujiwara2022hommage} for a recent survey of Chentsov's theorem whose precise formulation and proof are delicate when the state space is infinite.
\vspace{-10pt}
\subsection{CRLB and its variants}
The representation \eqref{eqn:KL.Hessian} of the Fisher information matrix $G(\theta)$ suggests that if $G(\theta)$ is ``large'', then it should be ``easier'' to distinguish, from data, between nearby distributions $\mathbb{P}_{\theta}$ and $\mathbb{P}_{\theta'}$. This intuition is made precise by the CRLB which provides a statistical interpretation of $G(\theta)$.

Suppose $\hat{\theta}$ is an {\it unbiased estimator} of $\theta$, i.e., $\hat{\theta}$ is a function of $X$ such that $\mathbb{E}_{\theta}[\hat{\theta}] = \theta$ for all $\theta$. The covariance matrix $\mathrm{Var}_{\theta}(\hat{\theta}) = (\mathrm{Cov}_{\theta}(\hat{\theta}_i, \hat{\theta}_j))_{i, j}$ of $\hat{\theta}$ quantifies the accuracy of $\hat{\theta}$. Under suitable regularity conditions, the basic Cram\'{e}r-Rao lower bound implies
\begin{equation} \label{eqn:CRLB}
\mathrm{Var}_{\theta}(\hat{\theta}) \geq G^{-1}(\theta),
\end{equation}
in the sense that the difference is positive semidefinite. In this context, the Fisher-Rao metric quantities the maximum information about $\theta$ one can gain from the data. Under suitable conditions, this lower bound is asymptotically attained by the {\it maximum likelihood estimator} as the sample size tends to infinity.

Interestingly, the Cram\'{e}r-Rao lower bound \eqref{eqn:CRLB} can be derived from an information geometric approach. Having defined the Riemannian metric by the formula \eqref{eqn:KL.Hessian}, one can derive \eqref{eqn:CRLB} using the Amari-Nagoaka theory \cite[Sec.~2.5]{amari2000methods}. Indeed, Eguchi \cite{eguchi1992geometry} defines a Riemannian metric from a divergence function using the formula \eqref{eqn:KL.Hessian}. This, when Amari-Nagoaka theory \cite[Sec.~2.5]{amari2000methods} is applied to it, has the potential to yield a new Cram\'{e}r-Rao type lower bound. In this way, \eqref{eqn:CRLB} has indeed been extended in many directions, some of which are summarized in Table \ref{tbl:summary}. For example, the Bayesian CRLB involves a positive function $\lambda(\theta)$ which plays the role of the prior density and relaxes the unbiasedness condition, and the $\alpha$-versions involve the {\it escort transformation} which transforms a density $p_{\theta}(x)$ to $p_{\theta}^{(\alpha)}(x) \propto p_{\theta}(x)^{\alpha}$. As shown in the table, these extensions require {\it modifications} of the Fisher-Rao metric. These modified Fisher-Rao metrics are obtained from different divergence functions. Some of these correspond to generalizations of the KL-divergence such as the R\'{e}nyi divergence \eqref{eqn:Renyi.divergence}, the $f$-divergence \eqref{eqn:f.divergence}, and the relative $\alpha$-entropy, together with the escort transformations; see \cite{mishra2021information} for details.

    \begin{table}
    \label{tbl:summary}       
    %
    %
    \centering
    \caption{CRLB variants and their Riemannian metrics}
    \begin{tabular}{cc}
    \hline\noalign{\smallskip}
    Bound  & Riemannian metric \\
    \noalign{\smallskip}
    \hline
    \noalign{\smallskip}
    Base CRLB \eqref{eqn:CRLB}   & $G(\theta)$ \\
    Bayesian and hybrid CRLB \cite{kumar2018information}  & $\lambda(\theta)\big[G(\theta) + J^{\lambda}(\theta)\big]$  \\
    Barankin bound \cite{barankin1949locally} & $G(\theta)$\\
    $\alpha$-CRLB \cite{kumar2020cram} &  $G^{(\alpha)}(\theta)$\\
    $(f, F)$-CRLB \cite{kumar2020cram} & $G^{(f, F)}(\theta)$\\
    Bayesian and hybrid $\alpha$-CRLB \cite{mishra2020generalized, mishra2021information} 
     & $\lambda(\theta)\big[G^{(\alpha)}(\theta) + J^{\lambda}(\theta)\big]$ \\
    \noalign{\smallskip}\hline\noalign{\smallskip}
    \end{tabular}
        
    \end{table}

\vspace{-10pt}
\subsection{Natural gradient learning and deep learning}
One of the most important consequences of the Fisher-Rao metric is the notion of {\it natural gradient} advocated by Amari \cite{amari1998natural}. Consider a function $f: \Theta \rightarrow \mathbb{R}$ defined in terms of a given parameterization of a statistical manifold $\mathcal{M} = \{ \mathbb{P}_{\theta} \}_{\theta \in \Theta}$. Let $G(\theta)$ be the Fisher information matrix. The {\it natural gradient} of $f$ at $\theta$ is given by
\begin{equation} \label{eqn:natural.gradient}
\nabla^G f(\theta) = G^{-1}(\theta) \nabla f(\theta),
\end{equation}
where $\nabla f$ denotes the usual Euclidean gradient. Geometrically, $\nabla^G f(\theta)$ corresponds to the Riemannian gradient with respect to the Fisher-Rao metric. Intuitively, the {\it natural gradient descent update}
\begin{equation} \label{eqn:natural.gradient.learning}
\theta^{(k + 1)} = \theta^{(k)} - \delta_k \nabla^G f(\theta^{(k)})
\end{equation}
follows the steepest descent direction under the Fisher-Rao geometry to minimize $f$. As shown by Amari, in statistical contexts natural gradient learning \eqref{eqn:natural.gradient.learning} leads to algorithms with asymptotically optimal convergence rates. Also see \cite{ollivier2018online} which interprets natural gradient learning in terms of a {\it Kalman filter}. When the underlying statistical manifold is an exponential family \eqref{eqn:exp.family}, (online) natural gradient learning is closely related to (online) mirror descent \cite{rm15}; this connection was recently extended in \cite{kainth2022conformal} to generalized exponential families using the $\lambda$-logarithmic divergence (see \eqref{eqn:log.divergence} below). 
\vspace{-10pt}
\subsection{Quantum divergence and its geometry}
In quantum metrology, the choice of measurement affects the probability distribution obtained. The implication of this effect is that the classical Fisher-Rao metric becomes a function of measurement. In general, there may not be any measurement to attain the resulting quantum Fisher-Rao metric  \cite{braunstein1994statistical}. There are many quantum versions of classical Fisher-Rao metric, e.g. based on the symmetric, left, and right derivatives \cite{yuen1973multiple} of the density operator that describes the quantum measurement of parameters.  Petz \cite{petz1996monotone,petz2007quantum} showed that all quantum Fisher-Rao metrics are a member of a family of Riemannian monotone metrics. Further, all quantum Fisher-Rao metrics yield quantum Cram\'{e}r-Rao inequalities with different achievabilities \cite{liu2019quantum}. For pure quantum states, the \textit{Fubini-Study metric} is a Riemannian metric that is directly linked to the quantum FIM. A similar relation for mixed states is available using the \textit{Bures metric}. Quantum algorithms to estimate von Neumann's entropy and $\alpha$-R\'{e}nyi entropy of quantum states (with Hartley, Shannon, and collision entropies as special cases for $\alpha=0$, $\alpha=1$, and $\alpha=2$, respectively) have also been reported  \cite{li2018quantum}. For the geometric structure induced from a quantum divergence, we refer the reader to \cite[Chapter 7]{amari2000methods}.
\vspace{-10pt}
\section{Radar Signal Processing}
An important engineering application of information geometry is in radar signal processing, where it has been applied to several tasks such as the measurement of information resolution \cite{cheng2012information,pribic2016stochastic2}, synthetic aperture radar (SAR) imaging \cite{gambini2014parameter,frery2012entropy}, and direction-of-arrival (DoA) estimation \cite{pribic2017information,coutino2016direction}.
\vspace{-10pt}
\subsection{Information resolution}
In a general radar sensing scenario, the radar sends a probing signal $\boldsymbol{\beta}$ toward the target. The measurement model of the signal reflected from the target toward the radar receiver is
\begin{align}
\mathbf{x} =h(\boldsymbol{\theta})+\mathbf{w},
\end{align}
where the function $h(\cdot)$ encapsulates the relationship between the measurement $\mathbf{x}$ and target parameter $\boldsymbol{\theta}$; and $\mathbf{w} \sim \mathcal{N}(0, \mathbf{N}(\boldsymbol{\beta}, \boldsymbol{\theta}))$
is the measurement noise that follows a zero-mean Gaussian distribution with covariance $\mathbf{N}(\boldsymbol{\beta}, \boldsymbol{\theta})$. The distance between two target states $\boldsymbol{\theta}_1$ and $\boldsymbol{\theta}_2$ can be measured in terms of geodesic distance $d(\cdot,\cdot)$ between two distributions $p\left(\mathbf{x} \mid \boldsymbol{\theta}_1\right)$ and $p\left(\mathbf{x} \mid \boldsymbol{\theta}_2\right)$ on the manifold $\mathbf{S}$ determined by the family of all such parameterized distributions. Then, all the equidistant points $\left\{\boldsymbol{\theta}^{\prime}\right\}$ with identical geodesic distance from $\boldsymbol{\theta}$ form an information resolution cell of $\boldsymbol{\theta}$ as
\begin{align}
\left\{\boldsymbol{\theta}^{\prime}\right\} \triangleq \left\{ \argmin _{\boldsymbol{\theta}^{\prime}} d\left(\boldsymbol{\theta}, \boldsymbol{\theta}^{\prime}\right)=\delta\right\}.
\end{align}
 
 The ability to distinguish two neighboring distributions $p\left(\mathbf{x} \mid \boldsymbol{\theta}^{\prime}\right)$ and $p(\mathbf{x} \mid \boldsymbol{\theta})$ at the radius $\delta$ establishes radar's resolution limit. The threshold $\delta$ for an information resolution cell is determined by defining tolerable probability of detection error. In this way, information geometry connects to the detection theory of radar.
\vspace{-10pt}
\subsection{SAR imaging}
SAR is a specialized radar system that employs a smaller antenna array but achieves the imaging performance of a large aperture through the movement of its platform. An interference pattern known as \textit{speckle} often poses challenges in retrieving the true SAR image from the measurement vector $\mathbf{z}$. For each $i$-th pixel, the received signal follows the multiplicative model $\mathbf{z_i}=\mathbf{x}_i \mathbf{y}_i$, where the speckle $\mathbf{y}$ is independent of the backscatter $\mathbf{x}$. The goal of an ideal speckle filter is to yield the estimator $\hat{\mathbf{x}}_i$ of the backscatter $\mathbf{x}_i$ using observations $Z$ such that the ratio $\mathbf{z}_i/\hat{\mathbf{x}}_i$ is a speckle: a collection of independent identically distributed samples from Gamma variates. Here, information-geometric distances are used to assess the quality of the filter by examining the closeness of the ratio image comprising $N$ pixels $\{\mathbf{z}_i/\hat{\mathbf{x}}_i\}_{i=1}^{N}$ to the hypothesis that it follows the (Gamma) distribution of a pure speckle \cite{gambini2014parameter}. The speckle filter may be optimized by analyzing these distances. 

In general, we choose one of the common distributions for SAR intensity return and then search for a distribution parameter, say $\widehat{\theta}$, in the parameter space $\Theta$ as
\begin{align}
\widehat{\theta}=\arg \min _{\theta \in \Theta} d_{h,\phi}(\widetilde{f}(\mathbf{z}), f(\mathbf{z }; \theta)),
\end{align}
where $\widetilde{f}$ and $f$ are the empirical and assumed distributions of $\mathbf{z}$, respectively; and $d_{h,\phi}(\cdot,\cdot)$ is generally one of the following \textit{$h$-$\phi$ distances} between the observations and the model: Bhattacharya, Rényi, triangular, harmonic, $\chi^2$, K-L, Hellinger, Havrda-Charvát, and Sharma-Mittal distances. These distances are defined by choosing suitable $h(\cdot)$ and $\phi(\cdot)$ functions in the expression:
\begin{align}
d_{h,\phi}\left(\theta_1, \theta_2\right)&=\frac{1}{2}\left(h\left(\int \phi\left(\frac{f\left(\theta_1\right)}{f\left(\theta_2\right)}\right) f\left(\theta_2\right)\right)\right. \nonumber\\
&\;\;\;\;\;\;\;\;\left.+ h\left(\int \phi\left(\frac{f\left(\theta_2\right)}{f\left(\theta_1\right)}\right) f\left(\theta_1\right)\right)\right),
\end{align}
which is the average of \textit{$h$-$\phi$ divergences} with a strictly increasing $h$, $h(0)=0$, and a convex, smooth $\phi$. 

Alternatively, Shannon geodesic distance may also be used \cite{naranjo2017geodesic}. In each case, the $h$-$\phi$ divergence or geodesic distance is turned into a test statistic, which may be used for various radar tasks such as edge detection, shape classification, estimation of filter weights, and measurement of the difference between samples. Similar to $h$-$\phi$ divergences, polarimetric SAR applications employ $h$-$\phi$ entropies that are also turned into test statistics with known asymptotic distributions. This approach is often used for edge detection with a small number of polarimetric data samples \cite{frery2012entropy}.
\vspace{-10pt}
\subsection{DoA estimation}
Some recent works have shown that common radar signal processing tasks such as beamforming can benefit from information geometric approaches. For example, consider a uniform linear antenna array with $M$ elements that admits signals from $K$ uncorrelated sources specified at time instant $l$ as $\mathbf{s}[l]=\left[s_1[l], \ldots, s_K[l]\right]^T$, each with zero mean and power $E\{s[l]s^H[l]\} = \sigma_s^2$, where $E\{\cdot\}$ denotes the mathematical expectation and $(\cdot)^H$ is the Hermitian operator. The sources impinge from directions $\boldsymbol{\theta}=\left[\theta_1, \ldots, \theta_K\right]^T$. The $M \times 1$ received signal vector $\mathbf{y}[l]=\left[y_1[l], \ldots, y_M[l]\right]^T$ for all antenna elements is
\begin{align}
\mathbf{y}[l] =\mathbf{A}(\boldsymbol{\theta}) \mathbf{s}[l]+\mathbf{n}[l],
\end{align}
where the noise vector $\mathbf{n}[l] \sim \mathcal{C N}\left(\mathbf{0}, \sigma_n^2 \mathbf{I}_M\right)$ is temporally and spatially white zero-mean Gaussian noise, $\textbf{I}_M$ is the identity matrix of dimension $M$, and $\mathbf{A}(\boldsymbol{\theta})=\left[\mathbf{a}\left(\theta_1\right), \ldots, \mathbf{a}\left(\theta_K\right)\right] \in \mathbb{C}^{M \times K}$ is the \textit{array manifold} matrix, whose $k$-th column is the steering vector 
\begin{align}
    \label{eq:stevec2}
    \textbf{a}(\theta_k)=[a_1(\theta_k),a_2(\theta_k),\dots, a_M(\theta_k)]^T,
    \end{align}
whose the $m$-th element is   
\begin{align}
    \label{eq:stevec1}
	a_m(\theta_k) = \exp \left\{\mathrm{j}\frac{2\pi d_m}{\lambda} \sin(\theta_k) \right\},
	\end{align}
where $d_m$ is related to the position vector of the $m$-th receive antenna, $\lambda$ is the operating wavelength, and  $\sin(\theta_k)$ is the \textit{DoA parameter}. The signal and noise are assumed to be stationary and ergodic over the observation period. 

The covariance matrix of the measurements is
	\begin{align}
    \label{eq:analytCovar}
	\textbf{R}_{yy} &= \text{E}\left\lbrace \textbf{y}\textbf{y}^H \right\rbrace = \sigma_s^2 \mathbf{A} \mathbf{A}^H + \sigma^2_n\textbf{I}_M,
	\end{align}
where $\textbf{I}_K$ is the identity matrix of dimension $K$. The DoA parameter is then estimated by solving the optimization problem
\begin{align}
\label{eq:doa}
\begin{array}{cc}
\underset{{\widetilde{\mathbf{R}}, \widetilde{\mathbf{A}} \in \mathcal{A}}}{\textrm{minimize}} & d\left(\mathbf{R}_{yy}, \widetilde{\mathbf{R}}\right) \\
\text {subject to } & \widetilde{\mathbf{R}}=\widetilde{\mathbf{A}} \widetilde{\mathbf{A}}^H,
\end{array}
\end{align}
where $\mathcal{A}$ is the set of feasible array manifold matrices based on the position vectors of array elements and number of sources; and $d(\cdot, \cdot)$ denotes a suitable information-geometric distance between distributions parametrized by the respective covariance matrices. The probability distribution $p(\mathbf{y} \mid \widetilde{\mathbf{R}})$ that is closest in information-geometric sense to the distribution described by the true covariance matrix. Conventional DoA estimation follows the least squares approach, where an Euclidean distance is minimized. However, \eqref{eq:doa} generalizes the minimization of error measure to an information-geometric distance. Note that this treatment is similar to the information-geometric generalization of least squares regression using Pythagorean theorem in Section~\ref{subsec:bregman}.
\vspace{-10pt}
\section{Optimal Transport} \label{sec:OT}
Optimal transport has become an extremely popular topic for information geometry in recent years. Here, 
we begin by specializing the optimal transport problem \eqref{eqn:Kantorovich} to the discrete case, not only to clarify the ideas but also to introduce without technical distractions some recent extensions. Thus let $\mathbb{P} = \sum_{i = 1}^n p_i \delta_{x_i}$ and $\mathbb{Q} = \sum_{j = 1}^m q_j \delta_{y_j}$ be discrete distributions on $\mathcal{X}$. A coupling $(X, Y)$ of $\mathbb{P}$ and $\mathbb{Q}$ is characterized by its joint probability mass function $\pi_{ij} = \mathrm{Pr}(X = x_i, Y = y_j)$. By construction, we have
\begin{equation} \label{eqn:pi.constraint}
p_i = \sum_{j = 1}^m \pi_{ij} \quad \text{and} \quad q_j = \sum_{i = 1}^n \pi_{ij}.
\end{equation}
Let $(c_{ij})$ be the cost matrix (possibly corresponding to a divergence on $\mathcal{X}$, i.e.,  $c_{ij} = D(x_i||y_j)$). Then \eqref{eqn:Kantorovich} is equivalent to a linear programming problem:
\begin{equation} \label{eqn:Kantorovich.discrete}
\min_{\pi} \sum_{i, j} c_{ij} \pi_{ij},
\end{equation}
over joint distributions $\pi$ satisfying \eqref{eqn:pi.constraint}. Intuitively, $(\frac{\pi_{ij}}{p_i})_j$ gives, for $i$ fixed, the conditional distribution for splitting the mass at $x_i$.

For both computation and modelling purposes it is often helpful to {\it convexify} the transport problem. By far the most important case is {\it entropic optimal transport} which regularizes \eqref{eqn:Kantorovich.discrete} by the (negative) Shannon entropy of $\pi$:
\begin{equation} \label{eqn:entropic.discrete}
\min_{\pi} \sum_{i, j} c_{ij} \pi_{ij} + \epsilon \sum_{i, j} \pi_{ij} \log \pi_{ij},
\end{equation}
where $\epsilon > 0$ is a tuning parameter. This problem can be solved quite efficiently using the {\it Sinkhorn algorithm} which amounts to iterative Bregman projections \cite{benamou2015iterative} and was recently shown to be an instance of mirror descent \cite{leger2021gradient, deb2023wasserstein}. The optimal value ${\bf D}_{\epsilon}(\mathbb{P}||\mathbb{Q})$ is often called the {\it Sinkhorn divergence} even though strictly speaking it is not a divergence (there are ways to modify it to yield a true divergence). 
For a broad overview of computational optimal transport and its applications see \cite{peyre2019computational}.

It is not difficult to imagine that the optimal transport problem \eqref{eqn:Kantorovich} or \eqref{eqn:Kantorovich.discrete} can be extended in various directions to suit specific needs. For example, we have {\it multi-marginal optimal transport} for dealing simultaneously with three or more distributions, {\it martingale optimal transport} for quantitative finance, and {\it quantum optimal transport} with quantum channels. Within information theory, there have been extensions which incorporate constraints on, say, the mutual information $I(X; Y)$, with applications in network information theory and rate distortion. A place to start is \cite{bai2023information}.

\vspace{-10pt}
\subsection{Wasserstein geometry}
To continue with our main theme we focus on the $2$-Wasserstein distance ${\bf W}_2$ on $\mathcal{P}(\mathbb{R}^n)$, which is the square root of the optimal transport cost when the ``ground divergence'' $D(x||y)$ is the square loss $|x - y|^2$ (see \eqref{eqn:W2}). For further details we recommend \cite{figalli2021invitation} which emphasizes the geometric aspects and is quite accessible as a first introduction to optimal transport. The $2$-Wasserstein distance is fundamental because:
\begin{itemize}
\item The square loss reflects the Euclidean geometry which is standard in many settings. 
\item The optimal transport can be characterized elegantly via convex duality. By {\it Brenier's theorem}, the optimal coupling in \eqref{eqn:W2} is deterministic and has the form $Y = \nabla \phi (X)$ for some {\it convex function} $\phi$. In particular, for distributions on the real line this reduces to a nondecreasing transformation. This characterization is useful not only for theoretical investigation but also numerical computation. For example, the $2$-Wasserstein transport between (multivariate) Gaussian distributions is known analytically and leads to the so-called {\it Bures-Wasserstein distance} between positive definite matrices. Also, {\it input convex neural networks} \cite{makkuva2020optimal}, among other methods, enable the use of machine learning algorithms for learning optimal transport maps using samples of $\mathbb{P}$ and $\mathbb{Q}$. 
\item The geometry induced on the space of probability distributions is tractable and has deep connections with analysis, probability, Riemannian geometry and partial differential equations.  
\end{itemize}
This {\it Wasserstein geometry} can be described via a {\it dynamic formulation} of the $2$-Wasserstein transport due to Benamou and Brenier. Instead of $\mathbb{P}$ and $\mathbb{Q}$, we denote the initial and terminal distributions by $\mathbb{P}_0$ and $\mathbb{P}_1$ respectively. Let $v_t(\cdot) : \mathbb{R}^n \rightarrow \mathbb{R}^n$ be a time varying vector field and let $\Phi_t(\cdot)$ be its flow, i.e., $x_t = \Phi_t(x_0)$ solves the ODE $\dot{x}_t = v_t(x_t)$ with initial value $x_0$. Let $X_0 \sim \mathbb{P}_0$ and define $X_t = \Phi_t(X_0)$, so that $\dot{X}_t = v_t(X_t)$. Consider minimization of the ``expected integrated kinetic energy'' given by
\begin{equation} \label{eqn:W2.dynamic.cost}
\mathbb{E} \Big[ \int_0^1 |\dot{X}_t|^2 \dd t \Big] = \mathbb{E} \Big[ \int_0^1 |v_t(X_t)|^2 \Big],
\end{equation}
subject to the terminal distributional constraint $X_1 \sim \mathbb{P}_1$. The flow along each vector field $v_t$ defines a coupling of $\mathbb{P}_0$ and $\mathbb{P}_1$. Then, it can be shown that the minimal value is ${\bf W}_2^2(\mathbb{P}_0, \mathbb{P}_1)$, and the optimal path has the form
\begin{equation} \label{eqn:McCann.interpolation}
X_t = X_0 + t (\nabla \phi(X_0) - X_0),
\end{equation}
where $\nabla \phi$ is the Brenier map. If we let $\mathbb{P}_t$ be the distribution of $X_t$, then the interpolation $(\mathbb{P}_t)_{0 \leq t \leq 1}$, which is called McCann's {\it displacement interpolation}, is a distance-minimizing geodesic under the $2$-Wasserste distance:
\[
{\bf W}_2(\mathbb{P}_s, \mathbb{P}_t) = |s - t| {\bf W}_2(\mathbb{P}_0, \mathbb{P}_1), \quad s, t \in [0, 1].
\]
Note that in \eqref{eqn:McCann.interpolation} each ``particle'' travels along a Euclidean geodesic (straight line). In fact, the dynamic formulation also provides a notion of orthogonality, or more generally inner product. Consider two paths $\mathbb{P}_t$ and $\mathbb{Q}_t$ satisfying $\mathbb{P}_0 = \mathbb{Q}_0 = \mu$ and driven respectively by gradient vector fields $u_t = \nabla \phi_t$ and $v_t = \nabla \psi_t$ (this restriction may be motivated by Brenier's theorem; for full details see \cite{ambrosio2005gradient}). Following Otto, we define the inner product between their initial velocities by
\begin{equation} \label{eqn:Otto.metric} 
\langle \dot{\mathbb{P}}_0, \dot{\mathbb{Q}}_0 \rangle = \int_{\mathbb{R}^d} u_0(x) \cdot v_0(x) \dd \mu(x).
\end{equation}
This inner product, together with the dynamic formulation, allows us to think of the Wasserstein space of probabiliy distributions as an infinite dimensional Riemannian manifold. A significant consequence of \eqref{eqn:Otto.metric} is the concept of {\it Wasserstein gradient flow}, where the gradient is defined in terms of the Otto metric \eqref{eqn:Otto.metric}, analogous to how Amari's natural gradient \eqref{eqn:natural.gradient} is defined via the Fisher-Rao metric. Specifically, the {\it heat equation} $\partial_t p_t = \Delta p_t$, where $p_t$ is a probability density, may be regarded as the Wasserstein gradient flow of the (negative) differential entropy $\int_{\mathbb{R}^d} p(x) \log p(x) \dd x$. Many of the above results hold in more general settings and form the basis of recent extensions. For example, the {\it Schr\"{o}dinger bridge problem}, which is closely related to the entropic transport problem \eqref{eqn:entropic.discrete} under the square loss, was given a dynamic formulation which amounts to augmenting \eqref{eqn:W2.dynamic.cost} with a regularization via the Fisher information of the interpolating densities \cite{chen2016relation}.
\vspace{-10pt}
\subsection{Relations between information geometry and optimal transport}
Given that both information geometry (in the traditional sense) and optimal transport study geometric structures on spaces of probability distributions, it is natural to study their relations and implications in applications. In this subsection we discuss some results related to the third author's research, and refer the reader to \cite{khan2022optimal} and the journal {\it Information Geometry} for further details and other developments. In particular, we mention \cite{ito2023geometric} which describes thermodynamic relations between information geometry and optimal transport, as well as \cite{li2023wasserstein} which develops a Wasserstein analogues of the Fisher information matrix \eqref{eqn:Hessian.as.Fisher.Rao} and the Cram\'{e}r-Rao inequality \eqref{eqn:CRLB}.

We begin by noting that the dualistic geometry in classical information geometry has a natural interpretation in terms of optimal transport \cite{WY19b}. While the precise mathematical formulation is beyond the scope of this article, we sketch the main ideas. 

Consider as a fundamental example the Bregman divergence \eqref{eqn:Bregman} which we rewrite, using convex duality, in the form (see \cite[Theorem 1.1]{amari2016information})
\begin{equation} \label{eqn:Bregman.self.dual}
D_{\phi}(x||x') = \phi(x) + \phi^*(y') - x \cdot y'.
\end{equation}
where $y' = \nabla \phi(x')$. Regard the mirror map $\nabla \phi$ as Brenier's optimal transport map for the $2$-Wasserstein transport. The expression \eqref{eqn:Bregman.self.dual}, which is nothing but the left hand side of the Fenchel-Young inequality \eqref{eqn:Fenchel.Young}, measures how much the point $(x, y')$ deviates from the graph of the optimal transport $y = \nabla \phi(x)$. It turns out that many divergences can be interpreted analogously by changing the cost function $c$ (ground divergence) of the transport problem and using a generalized $c$-convex conjugation; we call such a divergence a {\it $c$-divergence}. Moreover, the dualistic geometry of the $c$-divergence can be recovered using the geometry of optimal transport due to Kim and McCann \cite{kim2010continuity}. 

This viewpoint provides a systematic approach to generalize the Bregman divergence. Using the {\it logarithmic cost function} $c_{\lambda}(x, x') = \frac{-1}{\lambda} \log(1 + \lambda x \cdot x')$ (it recovers $-x \cdot x'$, which is equivalent to the square loss $\frac{1}{2}|x - x'|^2$, as $\lambda \rightarrow 0$) leads to the {\it logarithmic divergence} 
\begin{equation} \label{eqn:log.divergence}
\begin{split}
&D_{\lambda, \varphi}(x||x') = \phi(x) - \theta(x')\\
&\quad - \frac{1}{\lambda} \log(1 + \lambda \nabla \varphi(x') \cdot (x - x')),
\end{split}
\end{equation}
which is non-negative for  $\varphi$ such that $\frac{1}{\lambda}(e^{\lambda \varphi} - 1)$ is convex on a convex domain. Developed in \cite{w18, WZ21}, this logarithmic divergence has a tractable dualistic geometry and captures elegantly the R\'{e}nyi divergence \eqref{eqn:Renyi.divergence}, the $q$-exponential family mentioned earlier, as well as the ``diversification return'' in portfolio theory.

\begin{figure}[t!]
	\centering
	\vspace{-1.25cm}
	\includegraphics[width = 0.75\columnwidth]{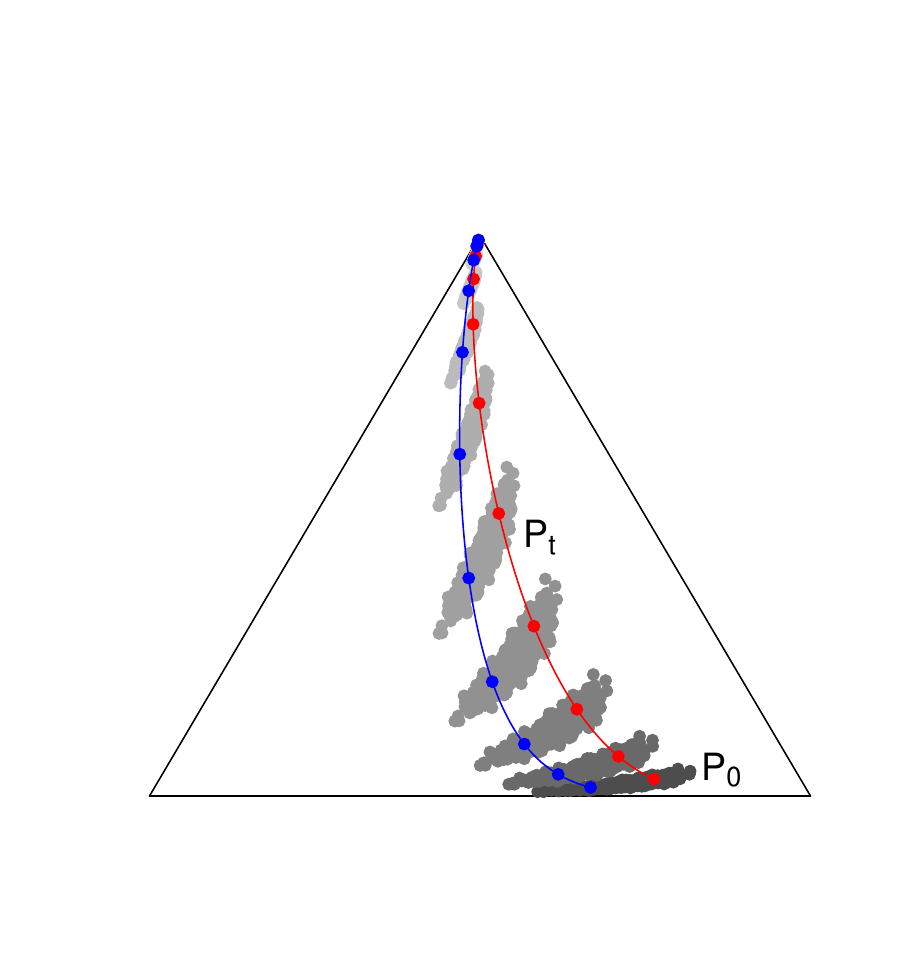}
    \vspace{-1.25cm}
	\caption{A Bregman-Wasserstein (primal) geodesic $(\mathbb{P}_t)_{0 \leq t \leq 1}$ on $\mathcal{P}(\Delta)$ constructed in \cite{rankin2023bregman}. Here each $\mathbb{P}_t$ is shown as a grey point cloud, and each ``particle'' in $\mathbb{P}_t$ travels along an exponential interpolation which is a geodesic with respect to the dualistic geometry induced by the KL-divergence on the simplex $\Delta$. The red and blue curves show the trajectories of two particles. This is analogous to, but different from, McCann's displacement interpolation \eqref{eqn:McCann.interpolation}.} \label{fig:BW}
\end{figure}

On the other hand, one can ask if a dualistic geometry can be directly constructed from an optimal transport divergence. In \cite{rankin2023bregman}, we showed that this is possible at least for the {\it Bregman-Wasserstein (BW) divergence} which is the optimal transport cost associated to a ground Bregman divergence cost (this is a special case of \eqref{eqn:Kantorovich}):
\begin{equation} \label{eqn:BW}
{\bf D}_{\phi}(\mathbb{P}||\mathbb{Q}) = \inf_{(X, X')} \mathbb{E} D_{\phi}(X||X').
\end{equation}
Letting $\phi(x) = |x|^2$ recovers the (squared) $2$-Wasserstein distance. To give another concrete example, consider the unit simplex $\Delta$ and equip it with the KL-divergence ${\bf H}(p||q) = \sum_i p_i \log \frac{p_i}{q_i}$ which is itself a Bregman divergence. Using \eqref{eqn:BW}, we may define, for $\mathbb{P}, \mathbb{Q} \in \mathcal{P}(\Delta)$ which may be regarded as distributions of {\it random} probability vectors, the BW-divergence given as the minimum expected KL-divergence of the optimal coupling. A ``primal geodesic'' under the induced generalized dualistic geometry is shown in Figure \ref{fig:BW}. Thanks to properties of the Bregman divergence, the BW-divergence provides a tractable extension -- both theoretically and computationally -- of the $2$-Wassertein distance. We claim, and hope to show in future research, that it will be useful in diverse applications including Wasserstein gradient flow, uncertainty quantification and distributional reinforcement learning.

\vspace{-10pt}
\section{Summary}
Geometric ideas have permeated the study and application of spaces of probability distributions. In this article, we provided an updated overview of key ideas of information geometry, as well as a glimpse of some recent research directions. We included a minimum of prerequisites and technicalities to enable an uninitiated information theorist to understand the breadth and depth of information geometry. Since an exhaustive treatment of all applications of interest is not possible here, we highlighted some ongoing research on information geometry in diverse applications such as estimation theory, learning theory, optimal transport, regression analysis, SAR imaging, and array signal processing based on the authors' own experience in the area.

 \bibliographystyle{IEEEtran}
 \scriptsize{\bibliography{main}}

\end{document}